# Investigation on the solar spectrum using Blackbody Radiation Inversion

Koustav Konar, Kingshuk Bose and R.K. Paul[*]

Department of Physics

Birla Institute of Technology

Mesra, Ranchi - 835215, Jharkhand

India

**Abstract**

The probability distribution of temperature of a blackbody can be determined from its power spectrum. This technique is called blackbody radiation inversion. In the present paper blackbody radiation inversion is applied on the spectrum of the Sun. The probability distribution of temperature and the mean temperature of the Sun are calculated without assuming a homogenous temperature and without using Stefan-Boltzmann law. Different properties of this distribution are characterized. This paper presents the very first mention and investigation of the distortions present within the Sun's spectrum.

**Introduction:**

Sun is the nearest star and the source of energy to us. We receive this energy in the form of radiation. The radiation can be described by treating the Sun as a blackbody radiator. Blackbody is an object capable of absorbing radiation of all frequency. And it emits radiation to maintain thermal equilibrium. This radiation is called blackbody radiation. The total power radiated per unit frequency per unit solid angle by a unit area of a blackbody emitter can be expressed by Planck's law [1,2]

$$P(\nu) = \frac{2h\nu^3}{c^2} \frac{1}{e^{\frac{h\nu}{kT}}-1} \qquad (1)$$

In eq. (1), $\nu$ is the frequency, T is the absolute temperature, h is Planck's constant, k is Boltzmann's constant and c is the speed of light. This law describes the spectrum for a blackbody that is in thermal equilibrium at a temperature T. But the temperature of the Sun is not homogenous and cannot be considered at thermal equilibrium at a single temperature. More appropriately the Sun is considered as a collection of blackbodies at local thermal equilibrium at a temperature T. Under this paradigm the Planck's law becomes [3]

$$W(\nu) = \frac{2h\nu^3}{c^2} \int_0^\infty \frac{\alpha(T)}{e^{\frac{h\nu}{kT}}-1} dT \qquad (2)$$

$$\text{Or, } G(\nu) = \int_0^\infty \frac{\alpha(T)}{e^{\frac{h\nu}{kT}}-1} dT \qquad (3)$$

Where $G(\nu) = \frac{c^2}{2h\nu^3} W(\nu)$. This step is incorporated for mathematical convenience.

In eq. (2), $W(\nu)$ is the radiated power per unit frequency per unit area and per unit solid angle and $\alpha(T)$ is the probability distribution of temperature of the blackbody. The dimension of $\alpha(T)$ is $\frac{1}{K}$.

Now, the blackbody radiation inversion (BRI) is to determine the probability distribution of temperature from the blackbody spectrum.

The solution of BRI requires solving eq. (3) which is a Fredholm integral equation of first kind and it is an ill-posed problem. The first attempt of solving the BRI was proposed by Bojarski by using Laplace transform with an iterative process [4]. Myriad of attempts to solve BRI have been made after this. Some of the methods present in the literature are Tikonov regularization method [5], universal function set method [6], Mellin transform method [7], modified Mobius inverse formula [8], variational expectation maximization method [9], maximum entropy method [10], regularised GMRES method [11]. Some of the other proposed methods are also available in the literature [12-15]. Recently a new method has been proposed in [16]. In the present paper the same is used.

For the spectrum of the Sun it is assumed that it has a temperature range of $T_1$ to $T_2$ and a frequency range of $\nu$. So, eq. (3) can be written as

$$G(\nu) = \int_{T_1}^{T_2} \frac{\alpha(T)}{e^{\frac{h\nu}{kT}}-1} dT \qquad (4)$$

The variable T is changed as $T = [T_1 + (T_2 - T_1)t]$. So, eq. (4) becomes [15]

$$G(\nu) = \int_0^1 K(\nu, t)\, a(t)\, dt \qquad (5)$$

Where $K(\nu, t) = \dfrac{T_2 - T_1}{e^{\left[\frac{h\nu}{k(T_1+(T_2-T_1)t)}\right]}-1} \qquad (6)$

And $a(t) = \alpha(T_1 + (T_2 - T_1)t)$.

The blackbody radiation inversion is to find out this $a(t)$ or $\alpha(T)$.

In this paper the data provided by the SOLSPEC spectrometer in the ATLAS and EUREKA missions are used [17]. And we have obtained the probability distribution of temperature of the Sun and the distortions present in its spectrum. In section 2, the method and its validation are described. And in section 3 the method is applied on the Sun's spectrum.

**Section 2: Method and Validation**

In this paper the method described in [16] is being used. And we are considering an analytical function, $a(t) = k_1 e^{-k_2 t^2} \sinh(k_3^2 t)$ as the solution of blackbody radiation inversion.

We have taken $T_1 = 4000$ K and $T_2 = 8000$ K such that $t = \dfrac{T-4000}{4000}$. This choice is made considering the temperature range of the sun.

Then, $\alpha(T) = k_1 e^{-k_2 \left(\frac{T-4000}{4000}\right)^2} \sinh\left(k_3^2 \dfrac{T-4000}{4000}\right) \qquad (7)$

These three unknown parameters $k_1$, $k_2$ and $k_3$ can be calculated by solving three integral equations similar to eq. (5) for three different frequencies. These equations can be obtained by transforming the spectral irradiance $I(\lambda)$ of the Sun's spectrum to the power spectrum $W(\nu)$ according to the relation [16]

$$W(\nu)d\nu = -I(\lambda)d\lambda \qquad (8)$$

This method is frequency dependent and its effectiveness has been validated for the frequency range of the cosmic microwave background spectrum ($\nu \sim 10^{11}$ Hz) in [16]. However, in the present application for the spectrum of the Sun the frequency range is different, $\nu \sim 10^{14}$ Hz. Hence, the method is validated for this frequency.

The reference data have been obtained by using the model function in eq. (9),

$$b(T) = e^{-\frac{(T-\delta)^2}{\gamma}} \qquad (9)$$

For figure 1 and 2, we have taken $b(T) = e^{-\frac{(T-5500)^2}{250000}}$. $b(T)$ is substituted in place of $\alpha(T)$ in eq. (4) to obtain the reference values for $G(\nu)$ for three different frequencies of $2\times10^{14}$ Hz, $3\times10^{14}$ Hz and $4\times10^{14}$ Hz. And $\alpha(T)$ (or, a(t)) is calculated by using these values of $G(\nu)$ in eq. (5) for the same frequencies [16]. The range of frequency is chosen to maintain the resemblance with the Sun's spectra. The model function and the reconstructed profile are plotted in figure 1. And the difference between $b(T)$ and $\alpha(T)$ is expressed as $d_1(T) = b(T) - \alpha(T)$. $d_1(T)$ is plotted in figure 2 against absolute temperature.

Figure 1

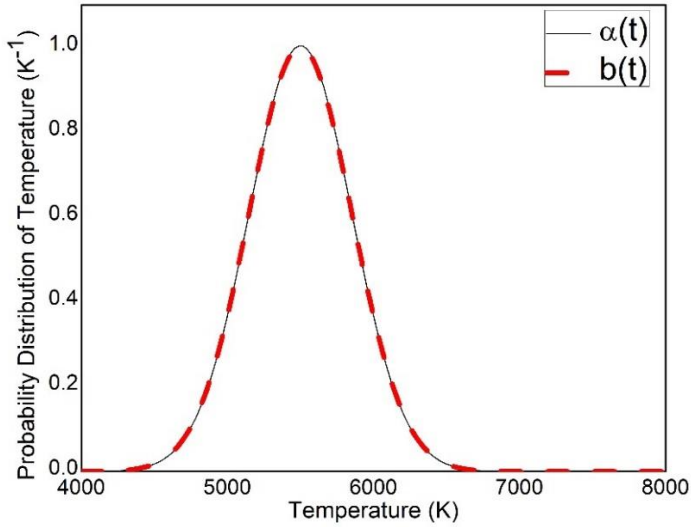

Figure 2

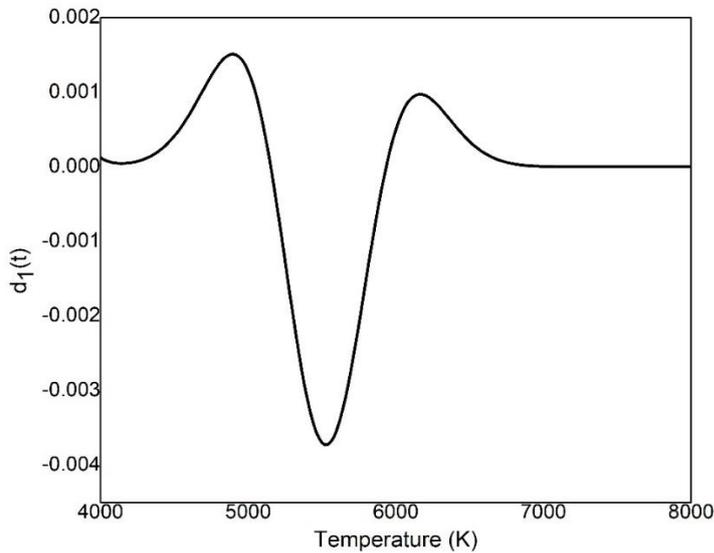

Figure 1: The model function $b(T)$ and reconstructed function $\alpha(T)$ are plotted against absolute temperature. Here $b(T) = e^{-\frac{(T-5500)^2}{250000}}$ and the reconstructed profile $\alpha(T)$ is obtained by using three frequencies of $2\times10^{14}$ Hz, $3\times10^{14}$ Hz and $4\times10^{14}$ Hz.

Figure 2: It shows the difference between the reference data b(t) and the reconstructed data $\alpha(T)$, $d_1(T) = b(T) - \alpha(T)$, plotted against absolute temperature.

The $|\frac{\Delta I}{I}|$ value obtained from figure 2 is $3.7\times10^{-3}$ for T = 5500 K, where I is the value of b(T) and ΔI is the value of $d_1(T)$. To verify this method in the frequency range of $\nu \sim 10^{14}$ Hz, we have taken sets of frequencies as i. Here i includes a set of three frequencies of $\nu_1 = i\times10^{14}$ Hz, $\nu_2 = (i+1)\times10^{14}$ Hz and $\nu_3 = (i+2)\times10^{14}$ Hz. This frequency sets are used to reconstruct the probability distribution of temperature, α(T). The standard deviation between the model data and reconstructed data is calculated using eq. (10)

$$\sigma = \sqrt{\frac{\sum_{j=1}^{N}[b_j(T)-\alpha_j(T)]^2}{N}} \quad (10)$$

where N is the number of data used for the calculation of standard deviation, N = 51 in our calculation. The standard deviation is plotted against chosen set of frequency.

Figure 3          Figure 4          Figure 5

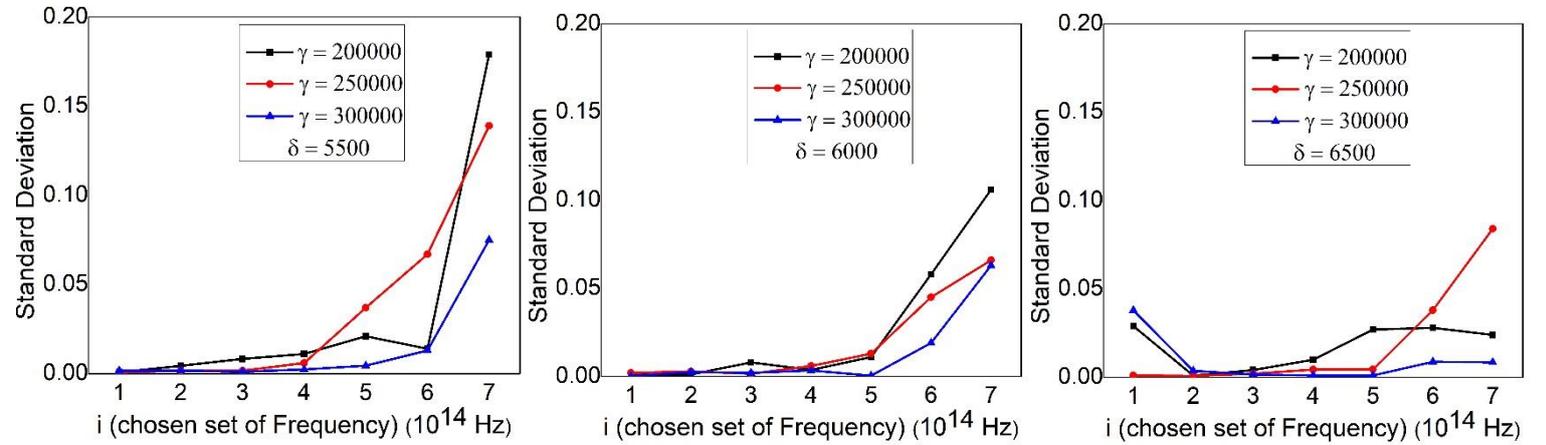

Figure 3, 4, 5: The standard deviation between the model data and reconstructed data is plotted against the chosen set of frequencies where each value of i includes a set of three frequencies of $\nu_1 = i\times10^{14}$ Hz, $\nu_2 = (i+1)\times10^{14}$ Hz and $\nu_3 = (i+2)\times10^{14}$ Hz used to calculate α(T). The model function b(T) is varied by changing δ and γ in eq. (10). The values of δ in figure 3 is 5500, in figure 4 δ = 6000 and in figure 5 δ = 6500.

The frequency sets of i = 2, 3 and 4 are preferred for minimum standard deviation. For a broader set of data, $\delta$ and $\gamma$ are varied in b(T) and α(T) is calculated each time. All the values of $k_1$, $k_2$ and $k_3$ that are calculated while changing b(T) are listed in fi 1, 2 and 3.

Table 1 ($\delta = 5500$)

| Frequency set ($\times 10^{14}$ Hz) | $\gamma = 200000$ | | | $\gamma = 250000$ | | | $\gamma = 300000$ | | |
|---|---|---|---|---|---|---|---|---|---|
| | $k_1$ | $k_2$ | $k_3$ | $k_1$ | $k_2$ | $k_3$ | $k_1$ | $k_2$ | $k_3$ |
| 1,2,3 | $2.64\times10^{-5}$ | 79.892 | 7.741 | $2.512\times10^{-4}$ | 63.87 | 6.921 | $1.097\times10^{-3}$ | 53.402 | 6.329 |
| 2,3,4 | $2.19\times10^{-5}$ | 81.236 | 7.807 | $2.337\times10^{-4}$ | 64.392 | 6.95 | $1.143\times10^{-3}$ | 53.101 | 6.31 |
| 3,4,5 | $3.868\times10^{-5}$ | 77.218 | 7.606 | $2.619\times10^{-4}$ | 63.59 | 6.905 | $1.081\times10^{-3}$ | 53.494 | 6.334 |

| 4,5,6 | 1.535×10⁻⁵ | 83.604 | 7.926 | 3.036×10⁻⁴ | 62.576 | 6.847 | 1.045×10⁻³ | 53.722 | 6.349 |
| 5,6,7 | 7.032×10⁻⁵ | 73.286 | 7.397 | 5.452×10⁻⁵ | 74.021 | 7.481 | 9.87×10⁻⁴ | 54.101 | 6.373 |
| 6,7,8 | 1.176×10⁻⁵ | 85.145 | 8.007 | 2.529×10⁻³ | 48.876 | 5.984 | 1.621×10⁻³ | 50.917 | 6.166 |
| 7,8,9 | 4.533×10⁻¹¹ | 165.236 | 11.319 | 1.935×10⁻⁷ | 108.543 | 9.183 | 7.351×10⁻⁵ | 70.06 | 7.334 |

Table 2 ($\delta = 6000$)

| Frequency set ($\times 10^{14}$ Hz) | $\gamma = 200000$ | | | $\gamma = 250000$ | | | $\gamma = 300000$ | | |
|---|---|---|---|---|---|---|---|---|---|
| | $k_1$ | $k_2$ | $k_3$ | $k_1$ | $k_2$ | $k_3$ | $k_1$ | $k_2$ | $k_3$ |
| 1,2,3 | 3.797×10⁻⁹ | 80.335 | 8.963 | 2.108×10⁻⁷ | 64.265 | 8.017 | 3.239×10⁻⁶ | 53.333 | 7.303 |
| 2,3,4 | 3.999×10⁻⁹ | 80.126 | 8.951 | 2.11×10⁻⁷ | 64.262 | 8.017 | 2.953×10⁻⁶ | 53.704 | 7.329 |
| 3,4,5 | 8.303×10⁻⁹ | 77.222 | 8.785 | 2.267×10⁻⁷ | 63.976 | 7.998 | 3.695×10⁻⁶ | 52.818 | 7.267 |
| 4,5,6 | 2.779×10⁻⁹ | 81.517 | 9.031 | 3.334×10⁻⁷ | 62.471 | 7.901 | 3.712×10⁻⁶ | 52.8 | 7.265 |
| 5,6,7 | 1.046×10⁻⁸ | 76.383 | 8.734 | 9.704×10⁻⁸ | 67.209 | 8.206 | 3.208×10⁻⁶ | 53.356 | 7.305 |
| 6,7,8 | 1.272×10⁻¹¹ | 101.978 | 10.138 | 3.725×10⁻⁶ | 53.388 | 7.274 | 1.08×10⁻⁶ | 57.42 | 7.592 |
| 7,8,9 | 1.101×10⁻⁵ | 50.164 | 6.991 | 1.294×10⁻⁹ | 82.922 | 9.165 | 7.681×10⁻⁵ | 41.72 | 6.402 |

Table 3 ($\delta = 6500$)

| Frequency set ($\times 10^{14}$ Hz) | $\gamma = 200000$ | | | $\gamma = 250000$ | | | $\gamma = 300000$ | | |
|---|---|---|---|---|---|---|---|---|---|
| | $k_1$ | $k_2$ | $k_3$ | $k_1$ | $k_2$ | $k_3$ | $k_1$ | $k_2$ | $k_3$ |
| 1,2,3 | 7.349×10⁻¹⁶ | 91.279 | 10.678 | 2.422×10⁻¹¹ | 64.354 | 8.969 | 3.904×10⁻⁸ | 45.397 | 7.527 |
| 2,3,4 | 5.489×10⁻¹⁴ | 79.937 | 9.996 | 2.77×10⁻¹¹ | 64.01 | 8.945 | 1.519×10⁻⁹ | 53.756 | 8.198 |
| 3,4,5 | 3.338×10⁻¹⁴ | 81.204 | 10.076 | 3.071×10⁻¹¹ | 63.746 | 8.926 | 2.088×10⁻⁹ | 52.947 | 8.135 |
| 4,5,6 | 2.406×10⁻¹³ | 76.213 | 9.757 | 2.083×10⁻¹¹ | 64.724 | 8.996 | 1.807×10⁻⁹ | 53.309 | 8.163 |

| | | | | | | | | |
|---|---|---|---|---|---|---|---|---|
| 5,6,7 | $1.34\times10^{-15}$ | 89.202 | 10.571 | $1.755\times10^{-11}$ | 65.151 | 9.026 | $9.168\times10^{-10}$ | 54.991 | 8.294 |
| 6,7,8 | $2.291\times10^{-12}$ | 70.711 | 9.386 | $1.083\times10^{-9}$ | 54.975 | 8.269 | $3.566\times10^{-9}$ | 51.666 | 8.031 |
| 7,8,9 | $1.624\times10^{-12}$ | 71.558 | 9.444 | $8.953\times10^{-16}$ | 88.992 | 10.602 | $3.798\times10^{-9}$ | 51.514 | 8.019 |

Table 1, 2, 3: It lists all the values of $k_1$, $k_2$ and $k_3$ of $\alpha(T)$ obtained while changing $b(T)$. All the values of $\delta$ and $\gamma$ are listed as well. In table 1 $\delta = 5500$, in table 2 $\delta = 6000$ and in table 3 $\delta = 6500$. The value of $\gamma$ is varied as 200000, 250000 and 300000 for each $\delta$.

### Section 3: Sun

Sun is the only star in our solar system. Studies on different features of the Sun have been carried out for millennium. One of the most obvious features to study is the temperature of the Sun. Generally, the Sun is treated as a blackbody and the radiation we receive is described by the Planck's law. The temperature of the Sun is calculated using Stefan-Boltzmann law. This process assumes that the photosphere is homogenous at a single temperature [18]. This also assumes an ideal gaussian distribution of temperature [19]. In this paper a more general method is proposed without these assumptions. It involves solving the blackbody radiation problem for Sun's spectra.

The blackbody radiation inversion (BRI), as discussed in [16], is applied to the spectrum of the Sun. We have used the data provided by the SOLSPEC spectrometer in the ATLAS and EUREKA missions for the spectrum of the Sun [17]. The temperature of the Sun is calculated by the application of BRI. The probability distribution of temperature and its properties are explained. This paper reveals the distortions present in the Sun's spectra for the first time. These distortions are caused by the mixing of different blackbodies at different temperatures [20].

Three frequencies and their corresponding intensities are used to calculate $k_1$, $k_2$ and $k_3$ of $\alpha(T)$ in eq. (7). We have listed the values of $k_1$, $k_2$ and $k_3$ in table 4.

Table 4

| | Frequency ($10^{14}$ Hz) | $k_1$ | $k_2$ | $k_3$ |
|---|---|---|---|---|
| $a_1(T)$ | 1, 2, 3 | $7.242\times10^{-6}$ | 4.239 | 0.145 |
| $a_2(T)$ | 1.5, 2.5, 3.5 | $7.07\times10^{-7}$ | 9.857 | 0.814 |
| $a_3(T)$ | 2, 3, 4 | $4.564\times10^{-3}$ | 10.963 | 0.01 |
| $a_4(T)$ | 2.5, 3.5, 4.5 | $1.626\times10^{-5}$ | 4.85 | -0.104 |
| $a_5(T)$ | 3, 4, 5 | $8.215\times10^{-6}$ | 5.934 | 0.169 |
| $a_6(T)$ | 3.5, 4.5, 5.5 | $1.511\times10^{-4}$ | 3.798 | -0.029 |
| $a_7(T)$ | 4, 5, 6 | $1.952\times10^{-5}$ | 4.845 | 0.094 |
| $a_8(T)$ | 4.5, 5.5, 6.5 | $1.299\times10^{-6}$ | 3.254 | 0.28 |
| $a_9(T)$ | 5, 6, 7 | $1.826\times10^{-3}$ | 11.65 | -0.02 |
| $a_{10}(T)$ | 5.5, 6.5, 7.5 | $1.916\times10^{-5}$ | 4.784 | 0.097 |

Table 4: $k_1$, $k_2$ and $k_3$ values for different probability distribution function are listed. The probability distributions are expressed as $a_1(T)$, $a_2(T)$, $a_3(T)$ and so on corresponding to a set of frequency.

M(T) is the average of all the probability distributions,

$$M(T) = \frac{\sum_{i=1}^{10} a_i(T)}{10} \qquad (11)$$

The average probability distribution M(T) is normalised with normalisation constant $1.217\times10^4$,

$$\alpha(T) = (1.217\times10^4) \times M(T) \qquad (12)$$

α(T) is the probability distribution of temperature of the Solar spectra.

In order to study different properties of the distribution, its moments of different orders are calculated using eq. (13)

$$n^{th} \text{ Moment} = \int_1^6 (T-\mu)^n \alpha(T) dT \qquad (13)$$

where n is the order of the moment and µ is the mean value.

First order moment or mean value is calculated as

$$\mu = \int_{4000}^{8000} T \times \alpha(T) dT \cong 5414 \qquad (14)$$

So, the mean temperature of the Sun is 5414 K.

Second order moment or variance is calculated as

$$\sigma^2 = \int_{4000}^{8000} (T-\mu)^2 \alpha(T) dT = 6.201\times10^5 \qquad (15)$$

So, standard deviation $\sigma = \sqrt{\sigma^2} \cong 788 \qquad (16)$

σ indicates the uncertainty in temperature in the distribution which is 788 K.

Third order standardised moment or Skewness is calculated as

$$\mu_3 = \int_{4000}^{8000} (T-\mu)^3 \alpha(T) dT = 3.451\times10^8 \qquad (17)$$

$$\beta_3 = \frac{\mu_3}{\sigma^3} = 0.706 \qquad (18)$$

In the calculation $\beta_3$ is a positive number. And a positive skewness describes the deviation from ideal gaussian behaviour, $\beta_3 = 0$ for ideal gaussian. This positive skewness arises due to the fact that the tail of the distribution right to the mean is more extended than the left-hand side tail which can be observed in figure 6 [21, 22].

$$\mu_4 = \int_1^6 (t-\mu)^4 \alpha(t) dT = 1.188\times10^{12} \qquad (19)$$

Excess kurtosis $\gamma_2 = \beta_2 - 3$, where $\beta_2 = \frac{\mu_4}{\sigma^4} = 3.09$. (20)

From eq. (20), $\gamma_2 = 0.09$ which is appositive number. Distribution with positive kurtosis is called Leptokurtic [20, 21].

To visualise the deviation from ideal gaussian behaviour, a Gaussian function ($\frac{1}{\sigma\sqrt{2\pi}} e^{-\frac{(x-\mu)^2}{2*\sigma^2}}$) is constructed in eq. (21).

$$s(T) = (5.04\times10^{-4}) \times e^{-\frac{(T-5414)^2}{1.24\times10^6}} \qquad (21)$$

The probability distribution of temperature α(T) and the gaussian distribution s(T) are plotted against absolute temperature in figure 6. The difference between α(T) and s(T), denoted by $d_2(T) = α(T) – s(T)$ is plotted against absolute temperature in figure 7.

Figure 6

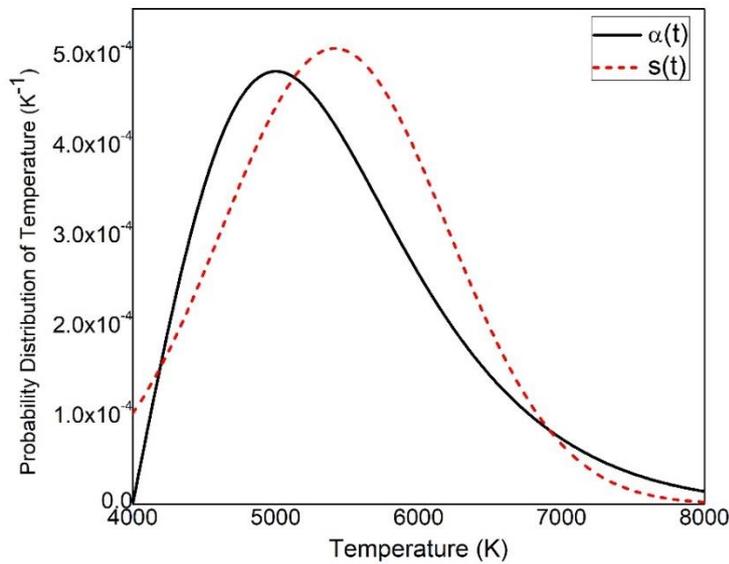

Figure 7

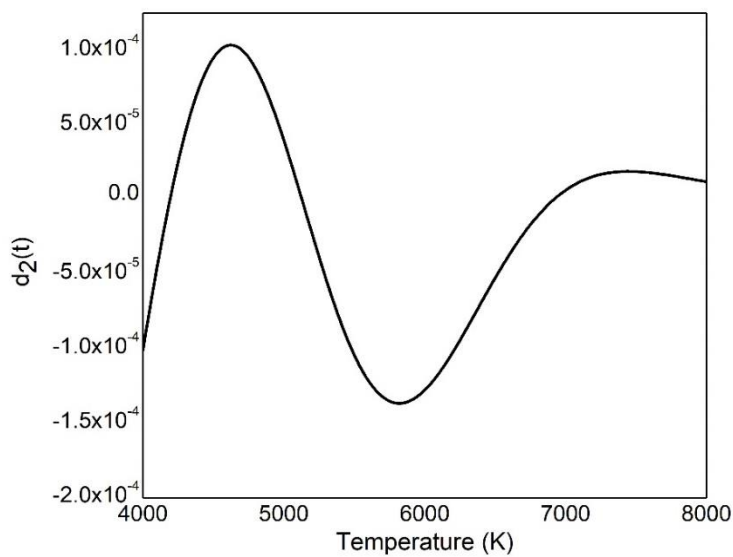

Figure 6: The calculated probability distribution of temperature α(T) and the Gaussian function s(T) are plotted against absolute temperature.

Figure 7: $d_2(T) = α(T) – s(T)$ and it is plotted against absolute temperature.

Figure 6 shows that the probability distribution of temperature of the Sun is not perfectly gaussian. A deviation is present. This deviation suggests that calculating the temperature by Stephan-Boltzmann law is not very accurate. A better approach is to use blackbody radiation inversion and determine the probability distribution of temperate. Then eq. (17) can be used to calculate the mean temperature. This temperature does not take the distortion we have found into account.

The calculated probability distribution of temperature should be able to reconstruct the intensity of the original input data. To verify this, the data used in the BRI are reproduced using the calculated probability distribution of temperature. Both sets of the intensities are plotted together against the frequency in figure 8. And all the values of the original data used as input and the reconstructed data are included in table 5.

Figure 8

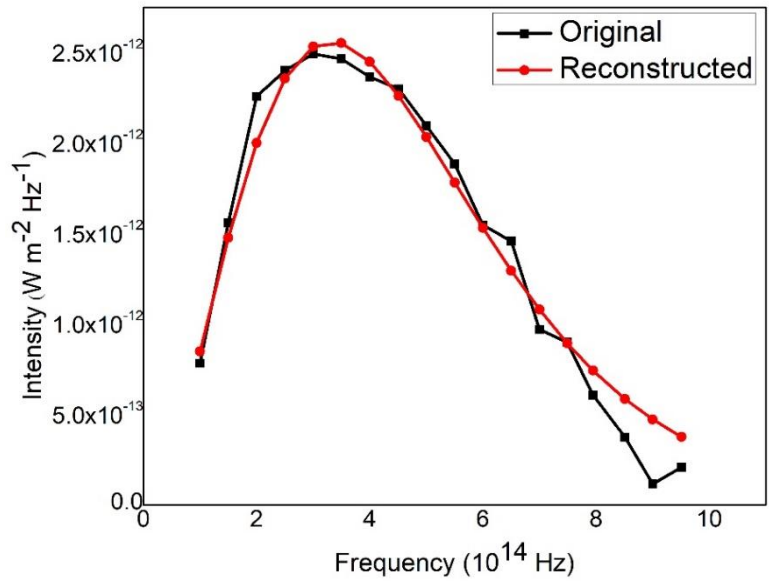

Figure 8: It plots the original input data and the reconstructed data together against frequency. The data are reconstructed by using the probability distribution of temperature we have obtained.

Table 5

| Frequency ($\times 10^{14}$ Hz) | Original data ($\times 10^{-12}$ W m$^{-2}$ Hz$^{-1}$) | Reconstructed data ($\times 10^{-12}$ W m$^{-2}$ Hz$^{-1}$) |
| --- | --- | --- |
| 1 | 0.7836 | 0.8498 |
| 1.5 | 1.56 | 1.477 |
| 2 | 2.258 | 2.001 |
| 2.5 | 2.401 | 2.357 |
| 3 | 2.493 | 2.534 |
| 3.497 | 2.466 | 2.553 |
| 4 | 2.366 | 2.45 |
| 4.511 | 2.3 | 2.261 |
| 5 | 2.095 | 2.034 |
| 5.5 | 1.886 | 1.783 |
| 6 | 1.548 | 1.532 |
| 6.5 | 1.459 | 1.296 |
| 7 | 0.9713 | 1.081 |
| 7.491 | 0.8993 | 0.8947 |
| 7.947 | 0.609 | 0.7443 |
| 8.511 | 0.376 | 0.587 |
| 9 | 0.1167 | 0.4743 |
| 9.509 | 0.21 | 0.3775 |

Table 5: All the original and reconstructed intensities are listed with corresponding frequencies.

In this paper the data collected by SOLSPEC spectrometer in the ATLAS and EUREKA missions are used [17]. These data represent the spectrum of the Sun as a black body radiator. After mathematically processing this data we have calculated the distortions present in the Sun's spectrum. The standard deviation between the original data and the reconstructed data is $0.143 \times 10^{-12}$ W m$^{-2}$ Hz$^{-1}$. This deviation represents the distortion present in the Sun's spectrum. Sun consists of different blackbody radiators of different temperatures ($\Delta T = \sigma$). The distortion in the spectrum we observe is the result of mixing of blackbodies with different temperatures [22, 23].

When this distortion is taken into account, the temperature of the Sun in our calculation is $T_{new} = T [1 + \langle (\frac{\Delta T}{T})^2 \rangle] \cong 5529$ K where T = 5414 K and $\Delta T$ = 788 K. The y and µ distortions, arising due to the mixing of blackbodies, are calculated as $y = \frac{1}{2} \langle (\frac{\Delta T}{T})^2 \rangle \approx 10^{-2}$ and $\mu = 2.8 \times \langle (\frac{\Delta T}{T})^2 \rangle \approx 10^{-2}$ [22, 23].

The processes responsible for the non-gaussian nature of the temperature distribution of the Sun is not fully understood yet. And future missions collecting data from the solar spectra and its coronal region plans to provide the necessary information to resolve the problem [23]. Parker Solar Probe which is already in its 6$^{th}$ orbit at the time of writing has provided critical information about the solar corona [24]. The Russian Interhelioprobe, consisting two interplanetary probes will provide the stereoscopic observation of solar activity [25].

**Discussion:**

Blackbody radiation inversion and its application on the spectrum of the Sun are studied. The probability distribution of temperature and the mean temperature of the Sun are calculated without assuming homogeneity of temperature. We have not used the Stefan-Boltzmann law either. The deviation of this probability distribution of temperature from ideal gaussian behaviour and its implications are analysed. The distortions present in the solar spectra due to mixing of blackbodies are delineated. This is the first study of these distortions in Sun's spectrum.

**Acknowledgement:**

We would like to thank the Department of Physics of Birla Institute of Technology, Mesra, Ranchi for furnishing a wonderful research environment. We also acknowledge and appreciate the help and support of B. Pathak, R. Kumar, M. K. Sinha (Department of Physics, BIT Mesra, Ranchi) and Soumen Karmakar (BIT Deoghar) *correspondence to: ratan_bit1@rediffmail.com